\documentclass{PoS}

\usepackage{xspace}
\newcommand{\Z}{\mbox{$Z$}\xspace}
\newcommand{\zjet}{\mbox{$Z${}-jet}\xspace}

\newcommand{\jt}{\mbox{$j_{\rm{T}}$}\xspace}

\newcommand{\pt}{\mbox{$p_{\rm T}$}\xspace}

\usepackage{lineno}
\title{Jet hadronization at LHCb}

\ShortTitle{Jet hadronization at LHCb}

\author{\speaker{Joseph D. Osborn}\\
        University of Michigan\\
        E-mail: \email{jdosbo@umich.edu}}

\abstract{In high energy proton-proton collisions, collimated sprays of particles, called jets, result from hard scattered quarks or gluons. Jets are copiously produced in these collisions; however, the dynamic process through which quarks and gluons, collectively referred to as partons, become bound state hadrons is still not well understood. Jets provide an excellent tool to study this process as they are proxies for the scattered parton; therefore, final-state hadrons can be measured with respect to an observable that is correlated to the scattered parton. The LHCb experiment is in an excellent position to measure hadrons within jets due to its excellent tracking and particle-identification capabilities. In this talk, new measurements of charged hadrons within jets measured opposite a \Z boson will be presented from the LHCb collaboration. }

\FullConference{13th International Workshop in High pT Physics in the RHIC and LHC Era (High-pT2019)\\
		19-22 March 2019\\
		Knoxville, Tennessee, USA}

\begin{document}

\section{Introduction}

Quantum chromodynamics (QCD) is unique amongst the four fundamental forces due to the nonperturbative dynamics that bind quarks and gluons within hadrons. Although quarks and gluons, referred to as partons, are the fundamental particles in QCD, they can never be observed freely due to the principle of confinement. The partonic structure of hadrons, with a focus on the proton, has been a deeply studied field of particle physics for decades. However, the process by which hadrons are formed from scattered partons is less understood than their initial-state partonic structure counterparts. Generally, fragmentation functions (FFs) are used to describe the probability of a particular parton fragmenting into a particular hadron~\cite{Collins:1981uk, Collins:1981uw}. Global fits to experimental data have improved FF phenomenology; for a recent review see Ref.~\cite{Metz:2016swz}. However, FFs are ultimately parameterizations of the underlying QCD interactions and represent a first approximation to understanding the mechanisms through which hadrons are formed from partons. Since hadronization is an inherently dynamic nonperturbative process and thus cannot be calculated perturbatively, additional data are required to further understand how hadrons form from scattered partons.

When a high-energy parton is scattered from a proton, the resulting collimated spray of particles is referred to as a jet. Jets are an excellent tool to study the hadronization process since they are a proxy for the scattered parton. Thus, unlike inclusive observables where only one hadron is measured in the final state, hadrons can be measured with respect to an object that is correlated with the fragmenting parton. The measurement of the jet allows for the transverse distribution of hadrons to be measured in addition to the more typical longitudinal momentum distribution of hadrons with respect to the scattered parton. Additionally, the jet provides an object which is calculable within perturbative QCD; thus, measurements of hadrons within jets utilize a final-state that is already perturbatively understood, the jet, to study something that is less understood and not perturbatively calculable, the result of the hadronization process.

In this talk, new measurements of charged hadron production within jets opposite a \Z boson are presented from the LHCb experiment~\cite{Aaij:2019ctd}. Previous measurements of hadrons within jets at the Large Hadron Collider (LHC) have been made almost exclusively in the inclusive jet channel~\cite{Aad:2011sc,Chatrchyan:2014ava,Acharya:2018eat}. At LHC energies, the inclusive jet channel is dominated by gluon jets, whereas the \zjet channel is predominantly sensitive to light quark jets~\cite{Sjostrand:2007gs}. Thus, these new measurements provide an opportunity to study phenomenological differences between light quark and gluon hadronization. A very recent measurement, studying the isolated photon-jet channel, is also dominated by quark jets and provides an opportunity to study hadronization differences between jets that are measured opposite a massive vs. massless vector boson~\cite{Aaboud:2019oac}. Additionally, previous measurements generally focus on the longitudinal momentum fraction of hadrons with respect to the jet, while the measurements presented here also study the charged hadron distributions transverse to the jet axis.

\section{Methods}

The methods used to construct charged hadron distributions within jets are similar to those in Ref.~\cite{Aad:2011sc}. Charged hadrons are measured as a function of three observables, and are normalized by the total number of \zjet pairs measured in a given jet transverse momentum (\pt) bin. This normalization ensures that the integral of the distributions gives the average multiplicity of the jets in a given jet \pt bin. Three charged hadron-in-jet observables are studied, defined as
\begin{equation}
    z\equiv\frac{{\bf{p}}_{\rm{jet}}\cdot{\bf{p}}_{\rm{hadron}}}{|{\bf{p}}_{\rm{jet}}|^2},
\end{equation}
\begin{equation}\label{eq:jt}
    \jt\equiv\frac{|{\bf{p}}_{\rm{jet}}\times{\bf{p}}_{\rm{hadron}}|}{|{\bf{p}}_{\rm{jet}}|},
\end{equation}
and
\begin{equation}\label{eq:r}
r\equiv\sqrt{(\phi_{\rm{jet}}-\phi_{\rm{hadron}})^2+(y_{\rm{jet}}-y_{\rm{hadron}})^2}.
\end{equation}
Here $\bf{p}$ is the 3-momentum vector, $\phi$ is the azimuthal angle, and $y$ is the rapidity. The observables give the longitudinal momentum fraction ($z$), the transverse momentum (\jt), and the radial distribution ($r$) of charged hadrons with respect to the jet axis.

The measurements are made by the LHCb experiment in $\sqrt{s}=8$ TeV proton-proton collisions. \Z bosons are measured via their dimuon decay. The dimuon pair is required to satisfy $60<M_{\mu^+\mu^-}<120$ GeV, where $M_{\mu^+\mu^-}$ is the invariant mass of the dimuon pair. Additionally, the muons must have $2<\eta<4.5$, where $\eta$ is the pseudorapidity. Jets are clustered with the $R=0.5$ anti-$k_T$ algorithm~\cite{Cacciari:2008gp} within $2.5<\eta<4$ and must have $\pt>20$ GeV. Requirements are placed on the \zjet pair to better identify events that correspond to a two-to-two partonic hard scattering process. Only events with one reconstructed primary vertex are analyzed, and the \zjet pair is required to be nearly back-to-back in azimuth such that $|\Delta\phi_{Z-{\rm{jet}}}|>7\pi/8$. Charged hadrons are required to have $|{\bf{p}}|>4$ GeV, $p_T>0.25$ GeV, and be within the jet such that $\Delta R \equiv \sqrt{(\phi_{\rm{jet}}-\phi_{\rm{hadron}})^2+(\eta_{\rm{jet}}-\eta_{\rm{hadron}})^2}<0.5$. Results are efficiency corrected and unfolded with a Bayesian unfolding procedure as implemented in Ref.~\cite{Adye:2011gm} to facilitate comparisons with theoretical calculations and predictions from event generators. The measurements are performed integrated over the \Z kinematics, so that the charged hadron-in-jet distributions can be studied in three bins of jet \pt.  Reference~\cite{Aaij:2019ctd} describes the analysis procedures and methods in greater detail.

\section{Results}

\begin{figure}[tbh]
	\centering
	\includegraphics[width=0.49\textwidth]{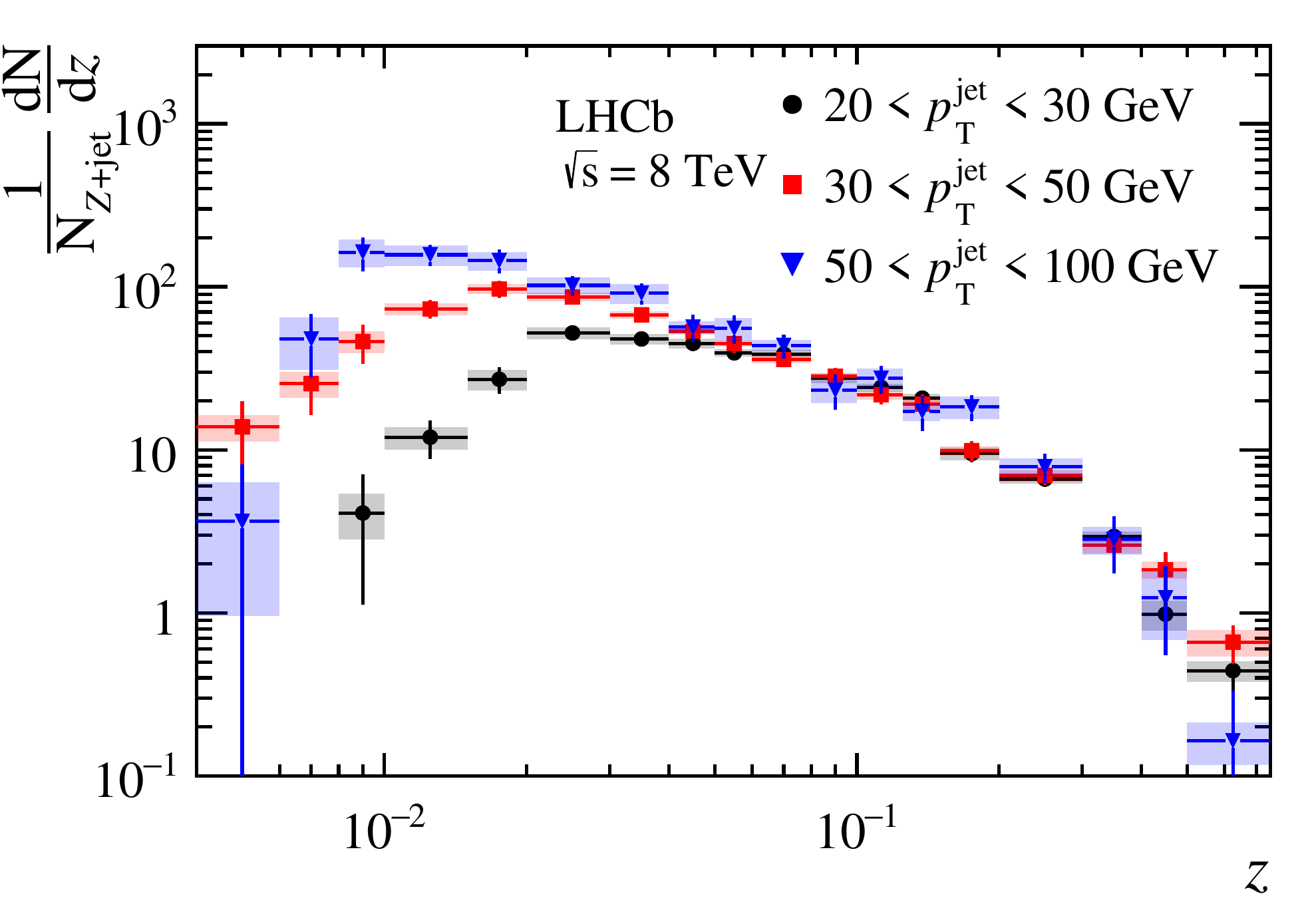}
	\caption{Measured longitudinal momentum fractions of charged hadrons in jets opposite a \Z boson.}
	\label{fig:z_dist}
\end{figure}

\begin{figure}[tbh]
	\centering
	\includegraphics[width=0.49\textwidth]{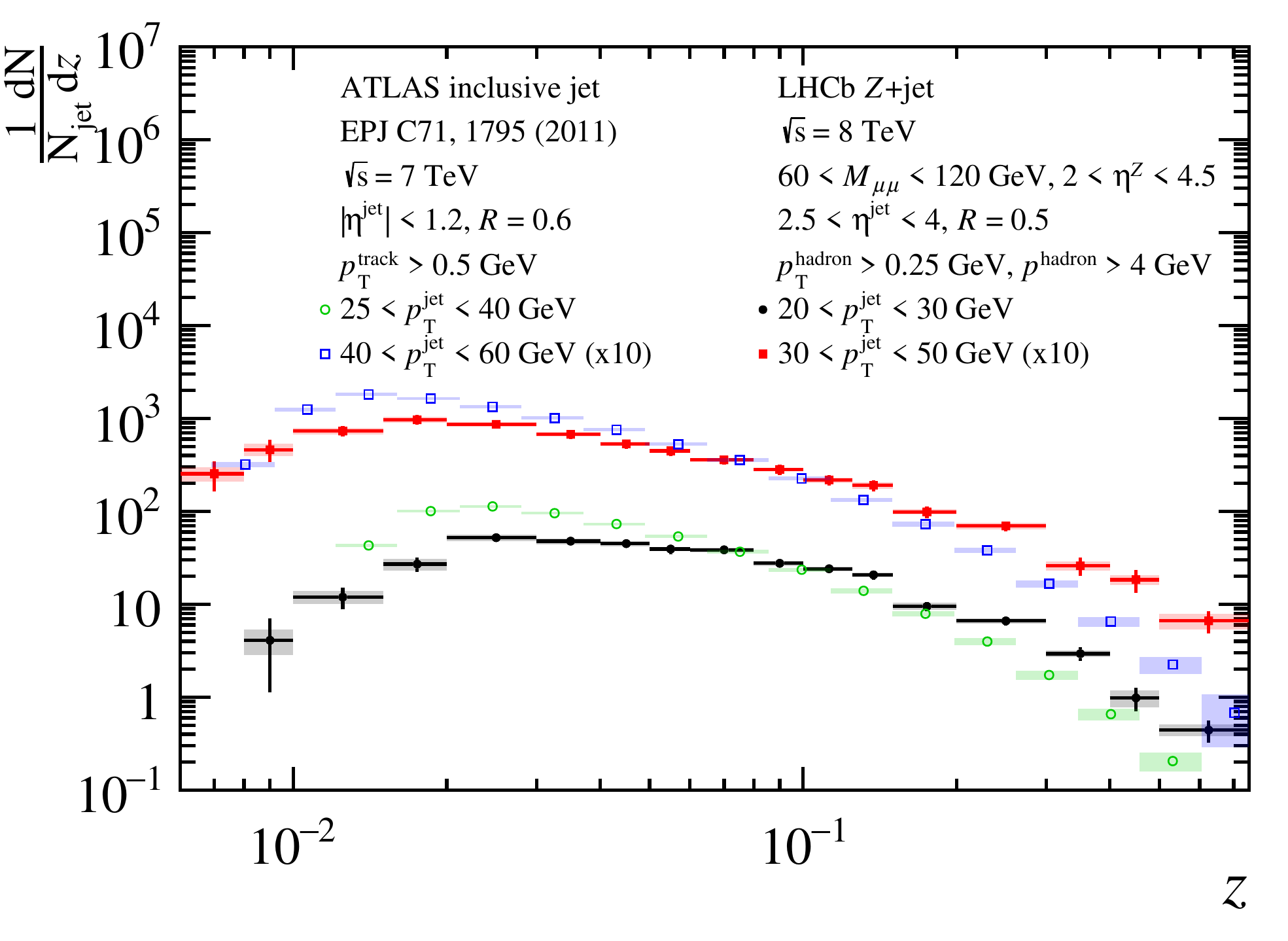}
	\includegraphics[width=0.49\textwidth]{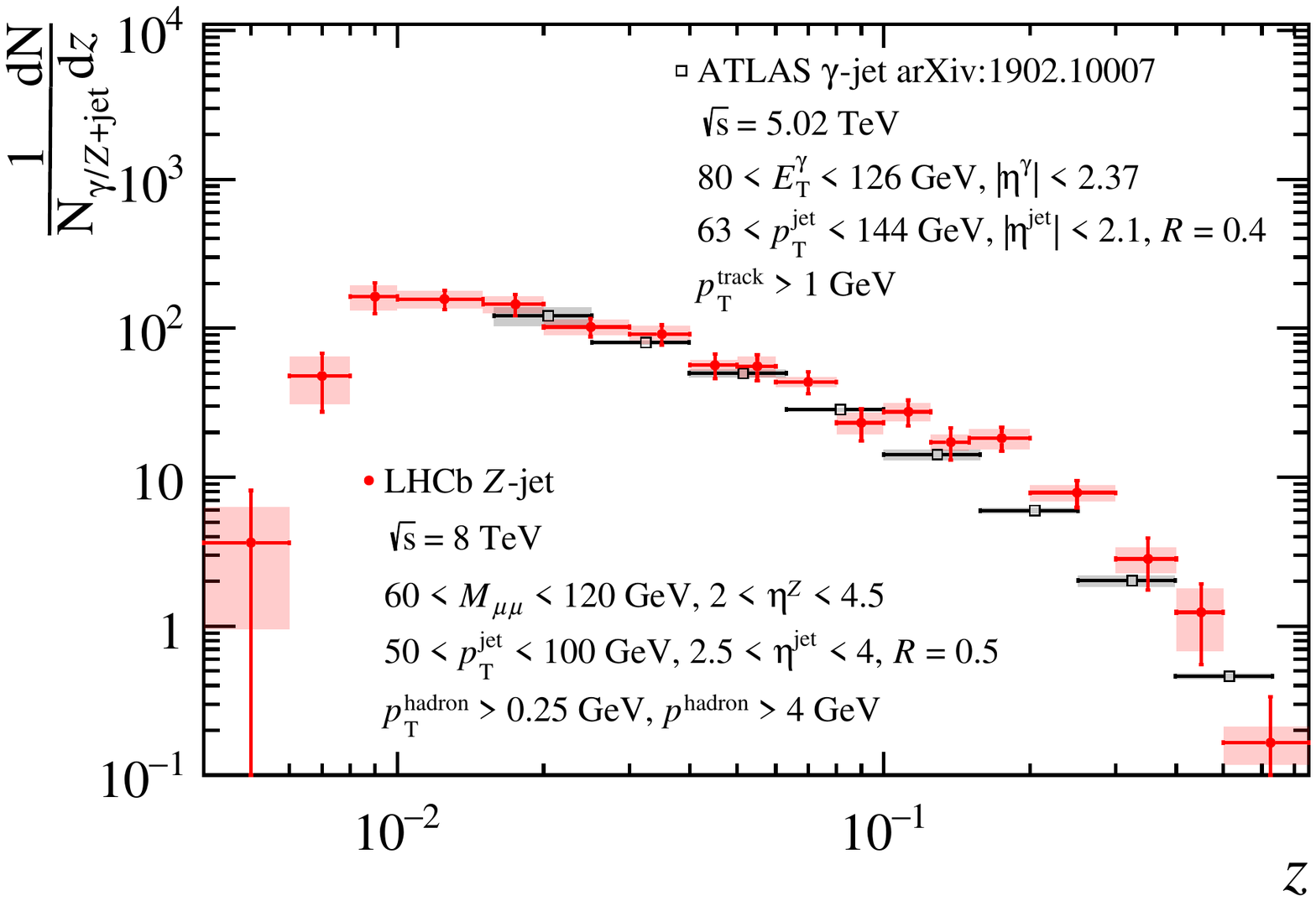}
	\caption{Comparisons of the longitudinal momentum fractions of charged hadrons in \zjet with similar measurements in midrapidity inclusive jets (left) and midrapidity isolated photon-jets (right).}
	\label{fig:z_comp}
\end{figure}

Figure~\ref{fig:z_dist} shows the charged hadron longitudinal momentum fractions in three jet \pt bins. The distributions of $z$ are relatively constant as a function of jet \pt at large $z$. At low $z$ the distributions diverge, which is a kinematic effect due to the requirement that the track momentum be greater than 4 GeV; therefore, higher \pt jets can probe lower $z$. Figure~\ref{fig:z_comp} shows comparisons of the longitudinal momentum fractions in the \zjet process to similar measurements in the midrapidity inclusive jet channel~\cite{Aad:2011sc} (left) and midrapidity isolated photon-jet channel~\cite{Aaboud:2019oac} (right). The comparison with inclusive jets shows that the fragmentation distributions are more steeply falling at high $z$ in the gluon dominated process when compared to the light-quark dominated \zjet process. Interestingly, the \zjet and isolated photon-jet distributions are extremely similar, indicating that, within the current uncertainties, the longitudinal hadronization is similar when the jet is measured opposite a massive \Z boson or a massless photon.

\begin{figure}[tbh]
	\centering
	\includegraphics[width=0.49\textwidth]{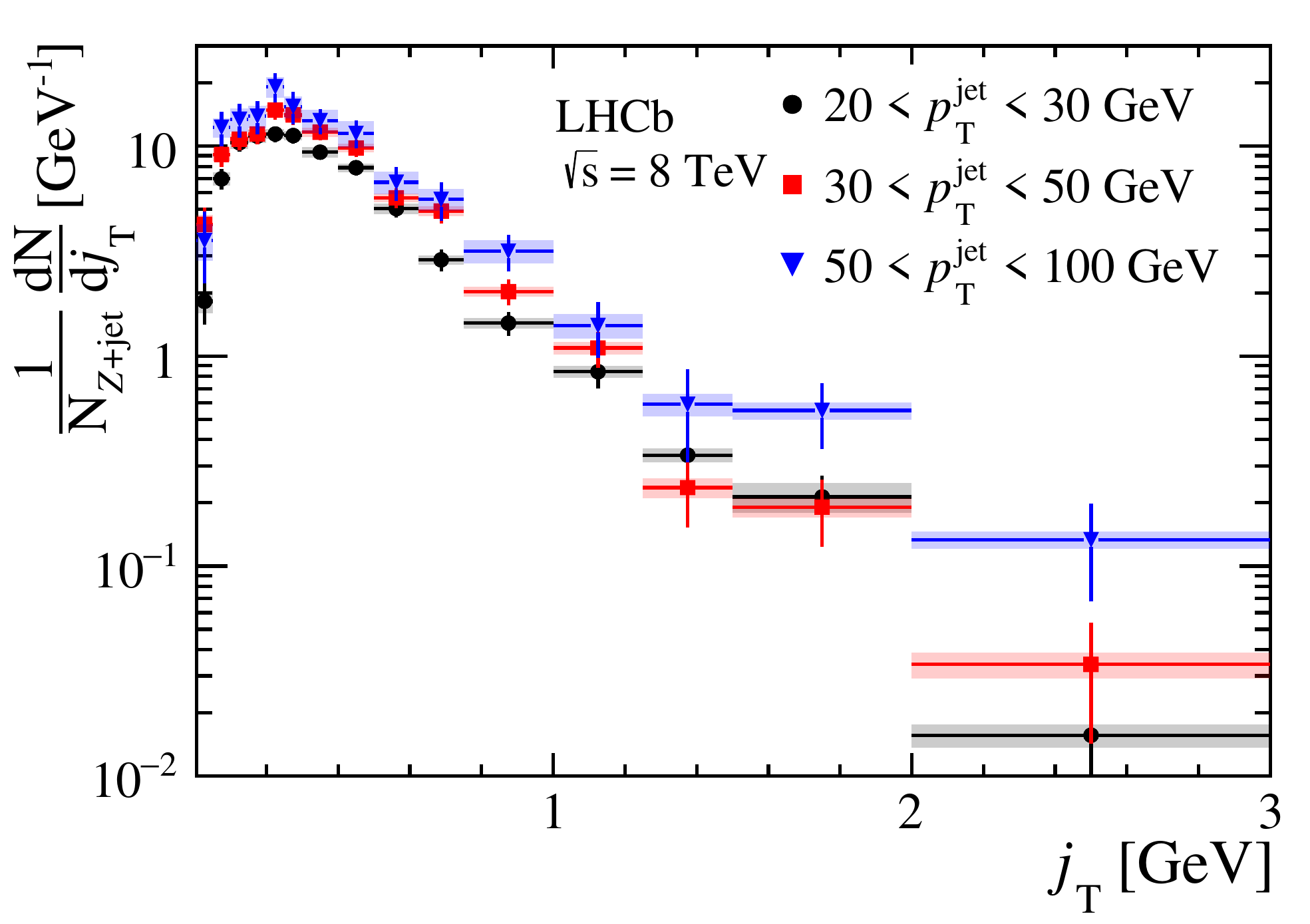}
	\includegraphics[width=0.49\textwidth]{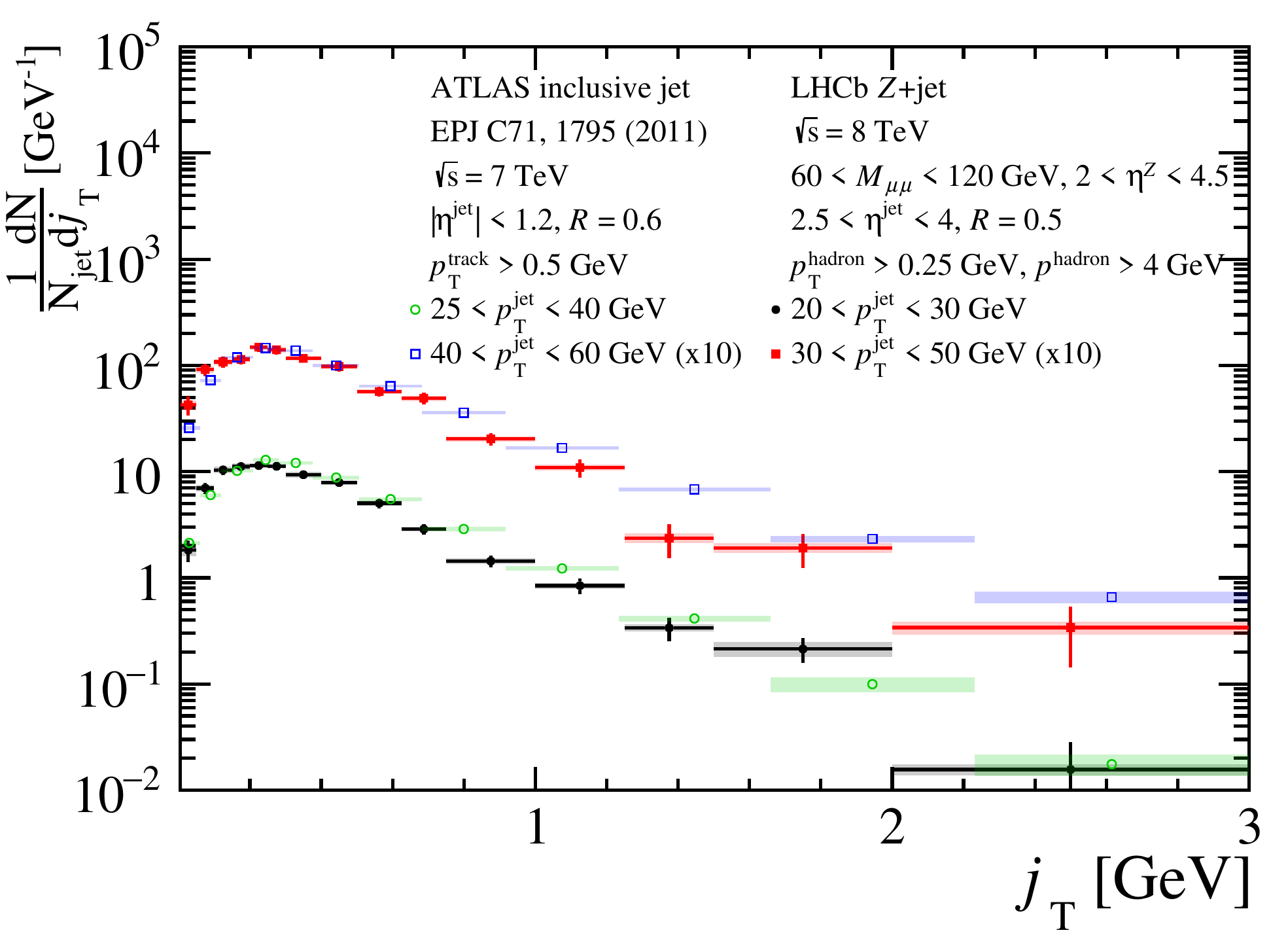}
	\caption{Measured transverse momentum with respect to the jet axis distributions as a function of jet \pt (left), and comparisons of these distributions to similar results from the midrapidity inclusive jet channel (right).}
	\label{fig:jt_dist}
\end{figure}

The charged hadron transverse momentum with respect to the jet axis distributions are shown in Fig.~\ref{fig:jt_dist} (left), while the comparison of these distributions with similar measurements in the midrapidity inclusive jet channel are also shown in Fig.~\ref{fig:jt_dist} (right). The distributions show a sensitivity to nonperturbative transverse momentum, indicated by the rounder shape at small \jt which transitions to a perturbative tail at large \jt. Performing a similar comparison to the midrapidity inclusive jet results shows that the distributions in the \zjet channel tend to lie at smaller \jt, which can be seen at $\jt<0.5$ GeV. At large \jt, the distributions are relatively similar; this may indicate that hadrons generated from perturbative radiation at large \jt are independent of the initial fragmenting parton.

\begin{figure}[tbh]
	\centering
	\includegraphics[width=0.49\textwidth]{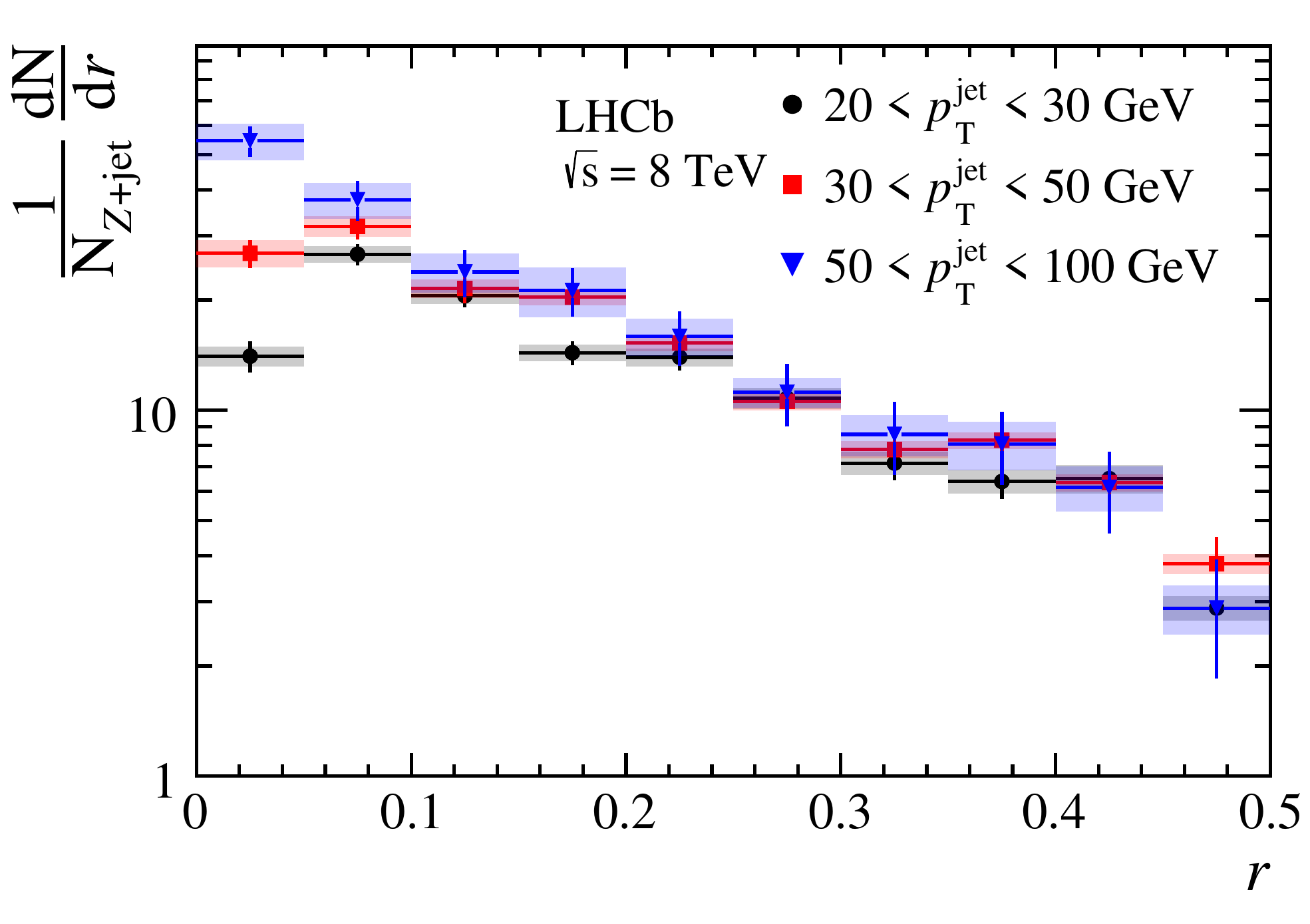}
	\includegraphics[width=0.49\textwidth]{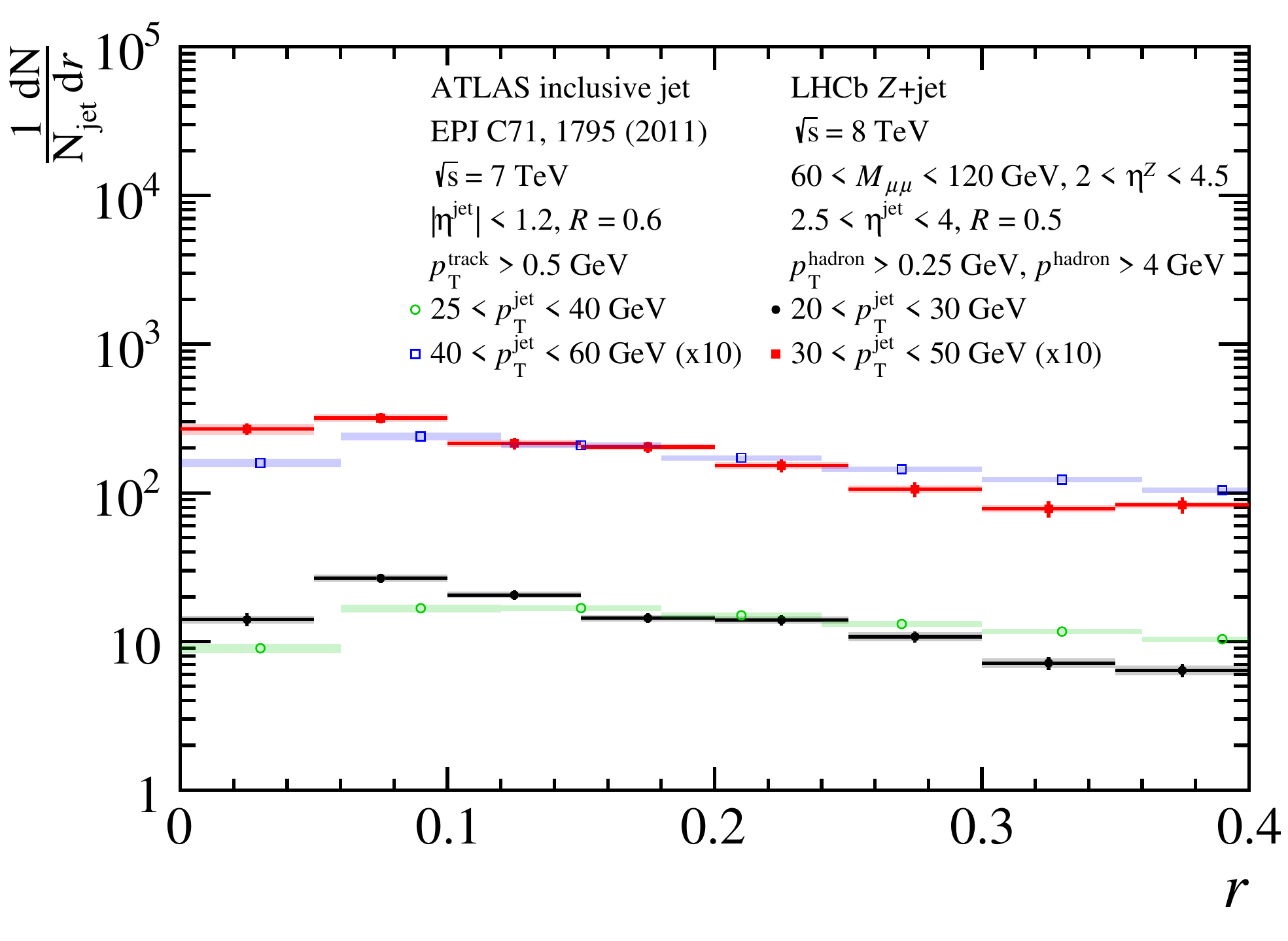}
	\caption{Measured radial distributions as a function of jet \pt (left), and comparisons of these distributions to similar results from the midrapidity inclusive jet channel (right).}
	\label{fig:r_dist}
\end{figure}

The charged hadron radial distributions as a function of jet \pt (left) and the comparison of these distributions to similar measurements in the midrapidity inclusive jet channel (right) are shown in Fig.~\ref{fig:r_dist}. The radial distributions are largely independent of jet \pt at large $r$, away from the jet axis. At small $r$, the multiplicity of hadrons rises sharply with the jet \pt which is a feature of the anti-k$_T$ algorithm. The independence of the distributions away from the jet axis may indicate that nonperturbative contributions to the jet are largely independent of jet \pt. Comparisons with the midrapidity inclusive jet data show that the radial distributions in the \zjet process are more collimated than those in the inclusive jet process. This same conclusion can also be drawn from the $z$ and \jt distributions, which display a more collimated behavior in the respective observable when compared to the midrapidity inclusive jet distributions.

In summary, new measurements of charged hadrons within jets measured opposite a \Z boson from the LHCb experiment are presented. These measurements are dominated by light quark jets, which is contrast to the majority of previous measurements from the LHC which are sensitive to predominantly gluon jets. Thus, the results offer the opportunity to study hadronization differences between light quarks and gluons. These measurements provide new information into the dynamic process of hadronization in both the longitudinal and transverse directions with respect to the jet axis. Future hadronization measurements from LHCb will utilize the excellent particle identification and heavy flavor jet identification capabilities already demonstrated by the experiment.

\bibliographystyle{ieeetr}
\bibliography{main}

\end{document}